\documentclass{article}
\usepackage{spconf,amsmath,graphicx}
\usepackage{booktabs}
\usepackage{amsfonts}
\usepackage{multirow}
\usepackage{bbding}
\usepackage{cite}
\usepackage[urlcolor=blue]{hyperref}
\usepackage{subfigure}
\usepackage{mathrsfs}
\usepackage{threeparttable}

\def\gcmodule{SubInter module}
\def\model{Inter-SubNet}
\title{Inter-SubNet: Speech Enhancement with Subband Interaction}
\name{
\begin{tabular}{@{}c@{}}
Jun Chen$^{1,2,\dagger}$\thanks{$^{\dagger}$ Work conducted when the first author was intern at Tencent.}, Wei Rao$^{2,*}$, Zilin Wang$^1$, Jiuxin Lin$^1$, Zhiyong Wu$^{1,3,*}$\thanks{$^{*}$ Corresponding author.}, \\\textit{Yannan Wang$^2$, Shidong Shang$^2$, Helen Meng$^{1,3}$}
\end{tabular}
}

\address{
  $^1$Shenzhen International Graduate School, Tsinghua University, Shenzhen, China\\
  $^2$Tencent Ethereal Audio Lab, Tencent, Shenzhen, China \\
  $^3$The Chinese University of Hong Kong, Hong Kong SAR, China \\
    \small{
        y-chen21@mails.tsinghua.edu.cn, ellenwrao@tencent.com, zywu@sz.tsinghua.edu.cn
    }
}
\begin{document}
\ninept
\maketitle
\begin{abstract}
Subband-based approaches process subbands in parallel through the model with shared parameters to learn the
commonality of local spectrums for noise reduction.
In this way, they have achieved remarkable results with fewer parameters.
However, in some complex environments, the lack of global spectral information has a negative impact on the performance of these subband-based approaches.
To this end, this paper introduces the subband interaction as a new way to complement the subband model with the global spectral information such as cross-band dependencies and global spectral patterns, and proposes a new lightweight single-channel speech enhancement framework called Interactive Subband Network (\model{}).
Experimental results on DNS Challenge - Interspeech 2021 dataset show that the proposed \model{} yields a significant improvement over the subband model and outperforms other state-of-the-art speech enhancement approaches, which demonstrate the effectiveness of subband interaction.
\end{abstract}
\begin{keywords}
% multi-scale time sensitive channel attention
subband interaction, global spectral information, \model{}, speech enhancement
\end{keywords}
\section{Introduction}
\label{sec:intro}

The interference from environmental noise is one of the main factors that hinder hearing aids, audio communication and automatic speech recognition.
% In these regards, single-channel speech enhancement methods, which remove background noise in the noisy audio and aim to improve the quality and intelligibility of speech, are highly desired.
In these regards, single-channel speech enhancement approaches, which eliminate interference noise in the noisy speech and aim to improve the clarity and comprehensibility of voice, are highly desired.
% , and have significant applications in hearing aids, audio communication and automatic speech recognition.
Traditional speech enhancement methods use statistical signal theory to effectively suppress stationary noise, but they do not perform well under conditions of low signal-to-noise ratio (SNR).
% Recently, with the aid of the powerful capabilities of deep neural networks, deep learning-based speech enhancement methods have yielded promising results, especially in dealing with non-stationary noise under challenging conditions, such as low SNR, reverberation, etc.
Recently, with the aid of the powerful capabilities of deep neural networks, deep learning-based speech enhancement approaches have yielded exceptional results, particularly in tackling challenging situations, such as low SNR, reverberation, etc.
% Noisy speech can be enhanced through neural networks either in time-domain or frequency-domain.
Noisy audio can be denoised through deep neural networks in the time or frequency domain.
The time-domain methods \cite{pandey2020densely, convtas, 2020unetgan, 2019tcnn, kong2022speech} generate the clean voice waveform straight from the noisy audio waveform.
Alternatively, the frequency-domain methods \cite{2014regression, wang2018supervised, tan2019learning, chen2017long} typically adopt the noisy spectrogram feature as their model's input feature, and the predicting target of them is the clean spectrogram feature or the spectrogram mask including Ideal Binary Mask (IBM) \cite{ibm}, Ideal Ratio Mask (IRM) \cite{irm} and complex Ideal Ratio Mask (cIRM) \cite{cirm}, just to name a few.
Overall, considering the robustness of system and computational complexity, the frequency-domain approaches are more preferable \cite{yin2020phasen}.

The subband-based approaches \cite{2020subband, lv2021dccrn+} are effective for single-channel frequency-domain speech enhancement.
This type of approaches divides the input spectrogram into multiple subbands and then processes them in parallel through the model with shared parameters, which helps the model learn the commonality of local spectrums and the frequency-wise signal stationarity to suppress noise\cite{2020subband}. 
Thus, these subband-based approaches reduce the number of parameters and computational burden while maintaining the model performance.

However, since they are only sensitive to the local information, the subband-based models can hardly recover the clean speech when encountering a complex acoustic environment with unclear local spectral patterns.
This challenge can be tackled with the help of the global spectral information, such as long-distance cross-band dependencies and global spectral pattern \cite{hao2021fullsubnet}: 
the model is capable of deducing the local pattern of the current subband from the pattern of the global spectrum as well as from other relatively clean subbands.
FullSubNet \cite{hao2021fullsubnet} and FullSubNet+ \cite{chen2022fullsubnet+} provide global spectral information to the subband model by serially connecting the fullband model with the subband model.
They work exceptionally well on the noise reduction task, which demonstrates the value of replenishing the global spectral information for the subband model.
Nevertheless, due to the introduction of the fullband model, these methods lead to a large number of overall model parameters as well as increase the complexity of the model.
% Nevertheless, due to the introduction of the fullband model, these methods lead to a large number of overall model parameters.
% Furthermore, it has been pointed out in \cite{chen2022speech} that this serial connection achieved by concatenation does not fully utilize the global information.

In this paper, to extract global spectral information like cross-band dependencies and global spectral patterns while maintaining a low model complexity, a novel subband interaction (SubInter) module is proposed for speech enhancement.
We incorporate the newly designed \gcmodule{} into the subband model \cite{2020subband} to complement the model with the global spectral information.
This new lightweight speech enhancement framework is called \model{},
which consists of alternating \gcmodule{} and Long Short-Term Memory (LSTM) layer to capture the global spectral information while retaining the ability to focus on the local spectral patterns.
Experimental results on DNS Challenge - Interspeech 2021 dataset show that our \model{} yields a dramatic improvement compared to the subband model.
Moreover, the \model{} outperforms FullSubNet+ and FullSubNet in terms of both number of parameters and performance, and also exceeds other state-of-the-art speech enhancement methods.
These results demonstrate that the subband interaction is a lightweight and efficient method for modeling the global spectral information. 
\begin{figure*}[!htbp]
	\centering
% 	\vspace{-0.4cm}
    \includegraphics[width=0.81\linewidth]{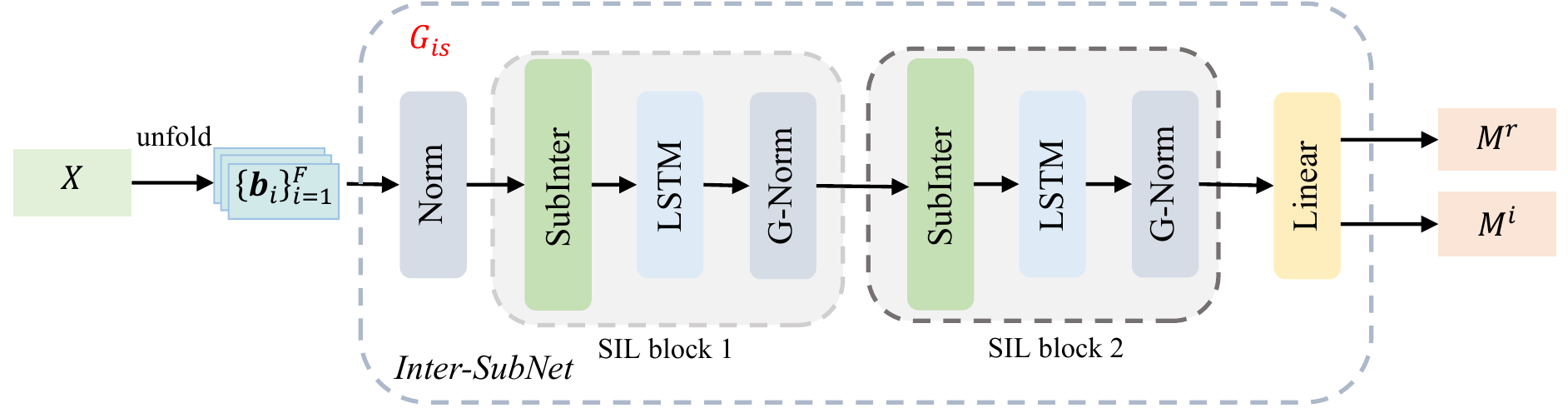}
    % \includegraphics[width=1.05\linewidth, height=0.6\linewidth]{src/architecture.pdf}
    % \vspace{-0.1cm}
	\caption{The general schematic of the proposed \model{}, where ``SIL block 1" and ``SIL block 2" refer to the first and second SubInter-LSTM block respectively. The ``G-norm" denotes group normalization \cite{gn}.
	The \model{} $G_{is}$ mainly consists of two stacked SIL blocks and one fully-connected layer.
	Taking the subband units $\{\mathbf{b}_i\}_{i=1}^F$ as input, the model $G_{is}$ generates the final output cIRM $\mathbf{M}^{r}$ and $\mathbf{M}^{i}$.}

	\label{fig:totalarch}
% 	\vspace{-0.1cm}
\end{figure*}

\section{Methodology}
\label{sec:format}

We consider single-channel signal with 16 kHz sampling rate after short-time fourier transform (STFT):
\begin{equation}
    % \mathbf{X} (t,f) = \mathbf{S}(t,f) + \mathbf{N} (t,f)
    % X (t,f) = S(t,f) + N (t,f)
    X_{t,f} = S_{t,f} + N_{t,f}.
\end{equation}
where $t=1,...,T$ and $f=0,...,F-1$ denote the frame and frequency indices, respectively. $X_{t,f}$, $S_{t,f}$ and $N_{t,f}$ are the complex-domain STFT factors of the noisy speech, noisy-free speech (the reverberant signal picked up by the microphone) and interference noise.
This paper only concerned with the denoising mission in the frequency-domain (STFT), and the goal is to filter the noise and recover the noisy-free speech signal with 16 kHz sampling rate.
% This paper focuses only on the denoising task in the short-time fourier transform (STFT) domain, and the target is to suppress noise and recover the speech signal (the reverberant signal received at the microphone) with 16 kHz sampling rate. 
% Towards this task, we propose a framework called \model{}, which extracts the global spectral information through \gcmodule{} while preserving the capacity of focusing on the local spectral patterns.

Towards this mission, we propose a framework called \model{}, which extracts the global spectral information such as cross-band dependencies and global spectral patterns through \gcmodule{}, while preserving the capacity of the subband-based model to concentrate on the local spectral patterns.
The overall diagram of \model{} $G_{is}$ is shown in Fig.\ref{fig:totalarch}. 
The input magnitude spectrogram $\mathbf{X} \in \mathbb{R}^{F \times T}$,
where $F$ and $T$ refer to the overall quantity of frequency bins and frames respectively,
is first transformed to subband units $\{\mathbf{b}_i\}_{i=1}^F$ by the ``unfold" procedure.
Taking the subband units as input, the model $G_{is}$ generates the final output cIRM $\mathbf{M}^{r}$ and $\mathbf{M}^{i}$.
The $G_{is}$ is mainly composed of two stacked SubInter-LSTM (SIL) blocks followed by one fully-connected layer.
We will elaborate each part in the following sub-sections.
% Next, we will go through each part in detail.

\subsection{Subband units}
\label{sec:subband units}
Subband units are obtained from magnitude spectrogram $\mathbf{X}$ through the ``unfold" operation and serve as the input to the \model{}.
Specifically, for each frequency $i$ from the magnitude spectrogram $\mathbf{X}$, we take a frequency bin vector $\mathbf{X}_{i} \in \mathbb{R}^{T}$ and the $2 \times n$ frequency bin vectors neighboring it as a subband unit $\mathbf{b}_{i}$:
\begin{equation}
\mathbf{b}_{i} = [\mathbf{X}_{i-n}, \cdots ,\mathbf{X}_{i}, \cdots , \mathbf{X}_{i+n}] \in \mathbb{R}^{F_s \times T}.
\end{equation}
where ``$[,]$" denotes the concatenation operation and $F_s = 2n + 1$. 
% In addition, circular fourier frequencies are used for boundary frequencies with $i+n > F-1$ or $i-n < 0$.
Besides, the boundary frequencies with $i+n > F-1$ or $i-n < 0$ are assigned circular fourier frequencies.
With the aforesaid process, we can ultimately acquire totally $F$ such subband units having a dimension of $(F_s, T)$, which are noted as $\{\mathbf{b}_i\}_{i=1}^F \in \mathbb{R}^{F \times F_s \times T}$.

\subsection{SubInter module}
In multi-channel speech separation, \cite{luo2020end} supplements global information with each channel through the interaction among the audio of distinct channels.
Inspired by this, we introduce \gcmodule{}, which exploits global spectral information including cross-band dependencies and global spectral patterns by the interaction among parallel subband units, and incorporates the global spectral information into these subband units.
In detail, following section \ref{sec:subband units}, we have a set of $F$ subband units $\{\mathbf{b}_i\}_{i=1}^F$ with $\mathbf{b}_i \in \mathbb{R}^{F_s \times T}$, where $F_s$ represents the number of frequency bins in a subband units.
The \gcmodule{} is applied on these subband units for complementary global spectral information:
\begin{equation}
\{\mathbf{\hat{b}}_i\}_{i=1}^F = \mathcal{H}(\{\mathbf{b}_i\}_{i=1}^F).
\end{equation}
where $\mathbf{\hat{b}}_i \in \mathbb{R}^{F_s \times T}$ is the subband feature that contains global spectral information as well as local spectral pattern, and $\mathcal{H}(\cdot)$ is the mapping function that is intended to describe the \gcmodule{}.

\begin{figure}[!htbp]
	\centering
% 	\vspace{-0.1cm}
    \includegraphics[width=0.58\linewidth]{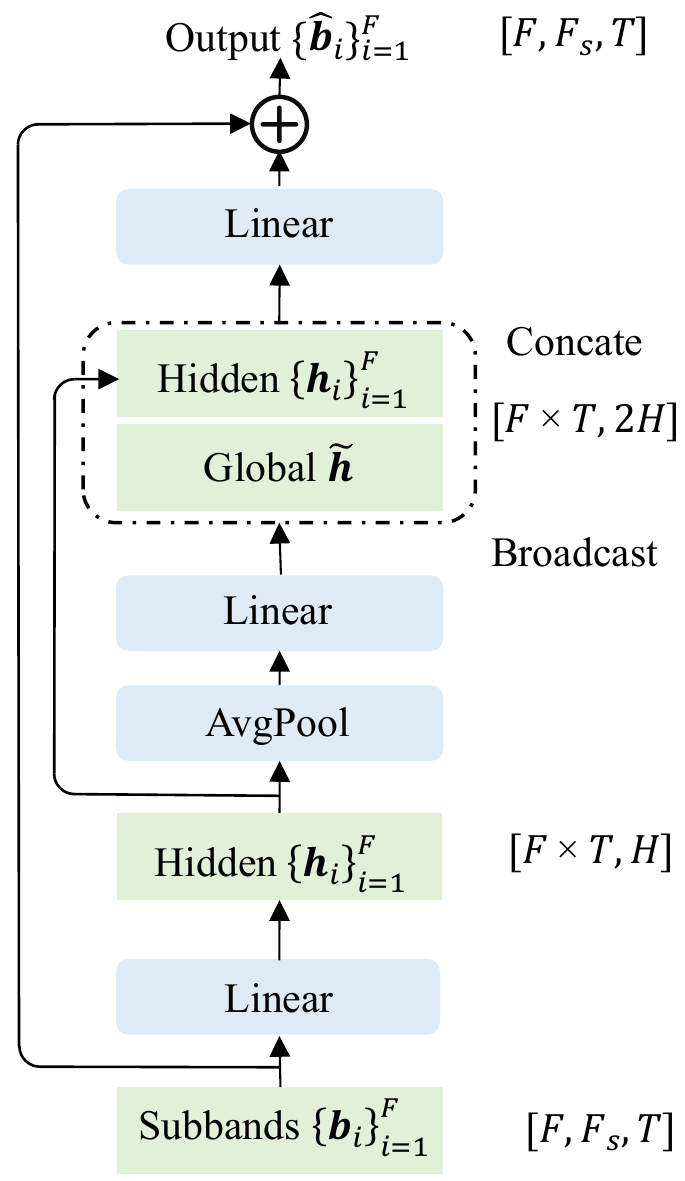}
    % \vspace{-0.1cm}
	\caption{The details of the SubInter Module. The ``Hidden" indicates the hidden representations. The ``Global" denotes the output of the second linear layer containing global spectral information.}
% 	\caption{The details of the SubInter module. The ``Hidden" indicates the hidden representations. The ``Global" denotes the output of the second linear layer with cross-band global information.}
	\label{fig:subinter}
% 	 \vspace{-0.1cm}
\end{figure}

With regard to the design of specific \gcmodule{}, as shown in Fig.\ref{fig:subinter}, for the set of subband units $\{\mathbf{b}_i\}_{i=1}^F$,  we transform the dimensions and then apply a linear layer with the mapping function $\mathcal{F}(\cdot)$ to obtain the hidden representation $\mathbf{h}_i \in \mathbb{R}^{T \times H}$:
\begin{equation}
\{\mathbf{h}_i\}_{i=1}^F = \mathcal{F}(transform(\{\mathbf{b}_i\}_{i=1}^F)).
\end{equation}
The $\{\mathbf{h}_i\}_{i=1}^F$ internally represent the information of parallel subband units.
Next, all of the $\mathbf{h}_i$s are averaged and sent to the second linear layer:
\begin{equation}
\mathbf{\tilde{h}} = \mathcal{R}(\frac{1}{F}\sum_{i=1}^F \mathbf{h}_i).
\end{equation}
where $\mathcal{R}(\cdot)$ represents the mapping function defined by the second linear layer.
This manipulation enables the otherwise parallel subband units to have an interaction, and hence the module extracts the $\mathbf{\tilde{h}} \in \mathbb{R}^{T \times H}$ containing the global spectral information as in cross-band dependencies and global spectral patterns.
Eventually, each $\mathbf{h}_i$ is concatenated with $\mathbf{\tilde{h}}$ to complement the global spectral information. 
And then they are fed into the third linear layer with the mapping function $\mathcal{P}(\cdot)$ for feature fusion. 
After the transformation of the dimensions and adding a residual connection from the module input $\mathbf{b}_i$, we get the output $\mathbf{\hat{b}}_i \in \mathbb{R}^{F_s \times T}$:
\begin{equation}
\{\mathbf{\hat{b}}_i \}_{i=1}^F = \{ \mathbf{b}_i\}_{i=1}^F + transform(\mathcal{P}(\{ [\mathbf{h}_i, \mathbf{\tilde{h}}] \}_{i=1}^F)).
\end{equation}
where ``$[,]$" is the concatenation operation.

Through the above-mentioned module, we extract the global spectral information by interacting the subband units that are originally processed in parallel, and then fuse the global information into the former subband units.
Consequently, the output subband features $\{\mathbf{\hat{b}}_i \}_{i=1}^F$ contain both global spectral information and local spectral pattern.

\subsection{Model architecture and workflow}
The original subband model is composed of two stacked LSTM layers, which process the subband units in parallel to model the signal stationarity and local spectral pattern.
To extract and integrate global spectral information, we propose the \model{}, which supplements the original subband model with \gcmodule{}.
In detail, SIL block is applied to replace the LSTM layer in original subband model.
As shown in Fig.\ref{fig:totalarch}, the \model{} $G_{is}$ mainly consists of two stacked SIL blocks and one fully-connected layer. \gcmodule{}, LSTM network and group normalization (G-norm) \cite{gn} layer are stacked in each SIL block.

According to section \ref{sec:subband units}, a total of $F$ subband units $\{\mathbf{b}_i\}_{i=1}^F$ are fed as the input to the $G_{is}$.
Inside SIL block, $\{\mathbf{b}_i\}_{i=1}^F$ are first interacted by the \gcmodule{}  to extract and fuse global spectral information to output the subband features $\{\mathbf{\hat{b}}_i \}_{i=1}^F$.
Then, the LSTM learns the local spectral patterns and the complementary global spectral information contained in these subband features.
After that, the G-norm layer normalizes the outputs of LSTM.
Finally, the fully-connected layer outputs the cIRM as the learning target.

% \section{Experiments}
% \label{sec:pagestyle}

\begin{table*}[!htbp]
    \begin{center}
    \caption{The performance in terms of WB-PESQ [MOS], NB-PESQ [MOS], STOI [\%], and SI-SDR [dB] on the DNS Challenge test dataset.} 
    % \vspace{-0.3cm}
    \label{table:large total compare}
    \scalebox{0.97}{
    \begin{tabular}{cccccccccccc}
    \toprule
        \multirow{2}{*}{Model}  & \multirow{2}{*}{Feat.} & \multirow{2}{*}{\shortstack{\# Para\\(M)}}&  \multicolumn{4}{c}{With Reverb} & \multicolumn{4}{c}{Without Reverb} \\
    \cmidrule(lr){4-7} \cmidrule(lr){8-11}
    &  &  &   WB-PESQ & NB-PESQ & STOI & SI-SDR & WB-PESQ & NB-PESQ & STOI & SI-SDR  \\
    \midrule
    Noisy  & - & - & 1.822 & 2.753 & 86.62 & 9.033 & 1.582 & 2.454 &  91.52 & 9.07 \\
    % \midrule
    DCCRN-E\cite{hu2020dccrn}  & RI &  3.70 & - & 3.077 & - & - & - & 3.266 & - & - \\
    Conv-TasNet\cite{convtas}  & Waveform & 5.08 & 2.750 & - & - & - & 2.730 & - & - & - \\
    PoCoNet\cite{isik2020poconet}  & RI & 50.0  & 2.832 & - & - & - & 2.748 & - & -  & - \\
    DCCRN+\cite{lv2021dccrn+}  & RI & 3.30 & - & 3.300 & - & - & - & 3.330 & - & - \\
    TRU-Net\cite{choi2021real} & Mag$^*$ & 0.38 & 2.740 & 3.350 & 91.29 & 14.87 & 2.860 & 3.360 & 96.32 & 17.55 \\    
    CTS-Net\cite{li2021two}  & Mag+RI  & 4.99 & 3.020 & 3.470 & 92.70 & 15.58 & 2.940 & 3.420 & 96.66 & 17.99 \\
    \midrule
    FullSubNet\cite{hao2021fullsubnet}  & Mag & 5.64 & 3.057 & 3.584 & 92.11 & 16.04 & 2.882 & 3.428 & 96.32 & 17.30 \\
    FullSubNet+\cite{chen2022fullsubnet+}  & Mag+RI  & 8.67 & 3.177 & 3.648 & 93.64 & 16.44 & \textbf{3.002} & 3.503 & \textbf{96.67} & 18.00 \\ 
    \midrule
    Subband model\cite{2020subband} & Mag & 1.82 & 2.885 & 3.475 & 92.10 & 15.54 & 2.552 & 3.242 & 95.10 & 16.46 \\
    Subband model$^L$  & Mag & 3.00 & 2.953 & 3.504 & 92.44 & 16.02 & 2.593 & 3.265 & 95.36 & 16.94 \\
    \model{}  & Mag & 2.29 & \textbf{3.207} & \textbf{3.659} & \textbf{93.98} & \textbf{16.76} & 2.997 & \textbf{3.504} & 96.61 & \textbf{18.05} \\
    \bottomrule
    \end{tabular}}
    \begin{tablenotes}
     \ninept{\item[1] $*$ More precisely, TRU-Net takes the per-channel energy normalization (PCEN) \cite{wang2017trainable} feature as input.}
   \end{tablenotes}
    \end{center}
    \vspace{-0.5cm}
\end{table*}

\section{Experimental Setup}

\subsection{Datasets}
The \model{} was trained on a sub-set of the DNS Challenge - Interspeech 2021 dataset with 16 kHz sampling rate, which will be called as DNS Challenge dataset in the following sections.
% To simulate actual acoustic scenarios, the DNS Challenge dataset is composed of a variety of clean speech, noise clips and room impulse responses (RIRs).
The clean speech dataset consists of 563 hours of clips from 2,150 speakers,  while there are 181 hours of over 60,000 clips from 150 classes of noise in the noise dataset.
To make the most of the dataset, dynamic mixing was utilized to mimic the noisy speech during model training.
% During model training, dynamic mixing was used to simulate speech-noise mixture as noisy speech to make full use of the dataset.
% In particular, 75\% of the clean speeches were mixed with the randomly selected room impulse responses (RIRs) from openSLR26 and openSLR28 \cite{ko2017study} datasets.
In particular, room impulse responses (RIRs), which are from the openSLR26 and openSLR28 \cite{ko2017study}, were randomly chosen and added to 75\% of the clean speech.
Afterwards, the noisy speeches were dynamically produced through mixing the clean or reverberant speeches and noise at a randomly selected SNR varying from -5 to 20 dB.
% The DNS Challenge also delivers a publicly available test set containing of two classes of synthetic clips \cite{reddy2020interspeech}, namely with and without reverberations.
The DNS Challenge also delivers a publically accessible test set containing synthesized clips of two classes \cite{reddy2020interspeech}, namely with and without reverberations.
There are 150 noisy clips with a SNR ranging from 0 to 20 dB in each class.
The effectiveness of the different models was verified on this test set.

\subsection{Training setup and baselines}
The Hanning window, which had 32 ms window length and 16 ms frame shift, was applied to convert the signal into the frequence-domain.
We employed Adam optimizer having a 1e-3 learning rate.
With respect to the subband units, $n=15$ was set by according to \cite{2020subband}, which means 15 neighbor frequencies are taken on each side of each input frequency bin.
% During training, the input-target sequence pairs are generated as constant-length sequences, with the sequence length set to $T = 192$ frames (approximately 3 s). 
During the training procedure, the input-target sequence pairs were generated into a constant length sequence of $T = 192$ frames (approximately 3 s). 

To testify the effectiveness of the subband interaction method, the following models were compared.
% For a fair comparison, all of the models used the same experimental settings as well as learning target (cIRM).
Aiming at a fair comparison, the identical experimental setup together with predictive objectives (cIRM) were implemented for all models.

\textbf{Subband model:} Following \cite{2020subband}, the subband model contained 2 LSTM layers and each layer consisted of 384 hidden cells.
The overall number of parameters of the subband model was 1.82 M.

\textbf{Subband model$^L$:} In order to verify that the performance improvement of \model{} is caused by the \gcmodule{} rather than the increase in the number of parameters, we designed a larger model named subband model$^L$.
By adding more LSTM layers, we increase the amount of parameters in the model.
The subband model$^L$, which comprised 3 LSTM layers with 384 hidden units for each layer, had a total of 3.00 M parameters.

\textbf{\model{}:} In the first SIL block, the \gcmodule{} had 102 hidden units, while the LSTM had 384 hidden units.
In the second block, there were \gcmodule{} containing 307 hidden units and the LSTM with 384 hidden units.
The number of parameters for \model{} was 2.29 M.

% \begin{table}[!htbp]
%     \begin{center}
%     \caption{Performance of WB-PESQ, and SI-SDR in the ablation study \ref{sec:ablation_study} using the test set without reverberation.}
%     % \vspace{-0.1cm}
%     \label{table:ablation}
%     \scalebox{1}{
%     \begin{tabular}{lcccc}
%     \toprule
%     Models  & WB-PESQ& $\Delta$WB-PESQ & SI-SDR & $\Delta$SI-SDR \\
%     \midrule
%     \model{}   & \textbf{3.282} & - & \textbf{17.20} & - \\    
%     \quad $-$ 2nd SubInter   & 3.208 & -0.074 & 16.56 & -0.64 \\
%     \qquad  $-$ 1st SubInter   & 3.177 & -0.031 & 16.44 & -0.12 \\
%     \bottomrule
%     \end{tabular}}
%     \end{center}
    % \vspace{-0.4cm}
% \end{table}

\section{Results and Discussions}

\subsection{Comparison with baseline and state-of-the-art methods}
Table \ref{table:large total compare} shows the performance of different speech enhancement models on the DNS Challenge dataset.
Over the table, ``With Reverb" and ``Without Reverb" mean the test sets with and without reverberation, respectively.
``\# Para" denotes the overall number of parameters of the model, which is measured in millions.
``Feat." stands for the input features of the model, where ``Waveform" refers to the time-domain waveform, ``Mag" indicates the frequency-domain magnitude spectrogram, and ``RI" indicates the frequency-domain real and imaginary spectrograms.
It should be noted that ``Mag+RI" means that all the magnitude, real and imaginary spectrograms are treated as input features by the model.

In the last three rows of Table \ref{table:large total compare}, the performances of subband model, subband model$^L$, and the proposed \model{} are compared and the results show that \model{} outperforms the subband model in all evaluation metrics.
It also shows that subband model$^L$ has improved performance compared to subband model after increasing the number of parameters.
However, when compared to subband model$^L$, the \model{} achieves superior results with fewer parameters, demonstrating that the performance of \model{} is improved by the \gcmodule{} rather than the increase in the number of parameters.

Additionally, from the table we can have a comparison of the \model{} with FullSubNet\cite{hao2021fullsubnet} and FullSubNet+\cite{chen2022fullsubnet+}.
It can be noticed that with much smaller number of parameters, our model outperforms FullSubNet in all metrics.
Moreover, compared with the more complex FullSubNet+, which takes all of magnitude, real and imaginary spectrograms as its input features, \model{} outperforms or equals the performance of FullSubNet+ using only the magnitude spectrogram as input.
These illustrate that, compared with serially connecting the fullband model with the subband model, our proposed \gcmodule{} is a more efficient and lightweight way for supplementing global spectral information.

We further compare the proposed \model{} to some other state-of-the-art time-domain and frequency-domain methods
\cite{hu2020dccrn, convtas, isik2020poconet, lv2021dccrn+, choi2021real, li2021two} 
on the DNS Challenge dataset
throughout Table \ref{table:large total compare}.
% It can be concluded that comparing with the latest methods, our proposed \model{} shows superior performance on noise reduction tasks without reverberation and even more prominent performance improvement with reverberation.
The conclusion can be drawn that, the \model{} shows superb performance in the denoising mission without reverberation compared to the latest methods, whereas the improvement in performance is even more prominent in the presence of reverberation.
These indicate that the proposed \model{} succeeds the remarkable capability of subband model in terms of reverberation effects that is reported in \cite{2020subband}, and drastically boosts the denoising capacity with \gcmodule{} for modeling global spectral information.

\subsection{Ablation study}
\label{sec:ablation_study}
In this section, ablation study was conducted to further explore the role of the \gcmodule{}.
Table \ref{table:ablation} presents the results of the ablation experiment, where ``1st SubInter" and ``2nd SubInter" refer to the \gcmodule{} in the first and the second SIL block, respectively.
``$\Delta$" represents the difference between the value after removing the module and the value before removing the module.
We can see that removing both the first \gcmodule{} and the second \gcmodule{} lead to a decline in the performance of the model in terms of signal (SI-SDR) and perceptual quality (WB-PESQ). 
This shows that subband interaction before each layer of subband processing LSTM network brings gains, and these gains can be superimposed without conflict.
In addition, the performance degradation caused by removing the first \gcmodule{} is greater than removing the second \gcmodule{}.
% 第一个SubInter是首次在Subband model补充全局信息，所以效果提升更大一些。
% This is probably because the first \gcmodule{} is the first one to introduce global spectral information into the subband model.
This is probably because that, without the second \gcmodule{}, the model still has the first \gcmodule{} to provide global spectral information;
whereas when the first \gcmodule{} is removed, the model degrades to the subband model lacking global spectral information, which does not perform well in complex acoustic environments.
In general, the usage of the \gcmodule{} does effectively improve the noise reduction capability of \model{}.

\begin{table}[t]
    \begin{center}
     \vspace{-0.2cm}
    \caption{WB-PESQ, $\Delta$WB-PESQ, SI-SDR and $\Delta$SI-SDR performance in the ablation study \ref{sec:ablation_study} on the without reverberation test set.}
    % \vspace{-0.2cm}
    \label{table:ablation}
    \scalebox{1.0}{
    \begin{tabular}{lcccc}
    \toprule
    \multirow{2}{*}{Models}  & \multirow{2}{*}{\shortstack{WB-\\PESQ}} & \multirow{2}{*}{\shortstack{$\Delta$WB-\\PESQ}} & \multirow{2}{*}{\shortstack{SI-\\SDR}} & \multirow{2}{*}{\shortstack{$\Delta$SI-\\SDR}}\\
     & & & & \\
    \midrule

    \model{}   & \textbf{2.997} & - & \textbf{18.05} & - \\    
    \quad $-$ 2nd SubInter   & 2.867 & -0.130 & 17.37 & -0.68 \\
    \qquad  $-$ 1st SubInter   & 2.552 & -0.315 & 16.46 & -0.91 \\
    \bottomrule
    \end{tabular}}
    \end{center}
    \vspace{-0.6cm}
\end{table}

\section{Conclusions}
\label{sec:conclusions}
This paper introduces a novel subband interaction (SubInter) module as a new way to model global spectral information for speech enhancement and proposes a new framework named \model{}\footnote{Code and examples: \href{https://github.com/RookieJunChen/Inter-SubNet}{https://github.com/RookieJunChen/Inter-SubNet}}.
This new lightweight speech enhancement framework is designed to exploit the global spectral information, meanwhile retaining the ability to focus on the local spectral patterns.
Experimental results demonstrate the effectiveness of the proposed \gcmodule{}. 
We also compare \model{} with other top-ranked methods on the DNS Challenge dataset, which shows that the superior performance of the proposed \model{}.
% The implementation of causal subband interaction will be the focus of our future work.

In the future, we will explore whether the subband interaction is applicable to other subband-based speech enhancement approaches. The implementation of causal subband interaction will also be the focus of our future work.

\textbf{Acknowledgement}: This work is supported by National Natural Science Foundation of China (62076144), Tencent AI Lab Rhino-Bird Focused Research Program (RBFR2022005) and Tsinghua University - Tencent Joint Laboratory.

\vfill\pagebreak

% \footnotesize
\bibliographystyle{IEEEbib}
\bibliography{strings,refs}

\end{document}